\documentclass[prb,twocolumn,showpacs,floatfix,amsmath,amssymb,superscriptaddress]{revtex4}
\usepackage[dvips]{graphicx}
\usepackage{latexsym}
\usepackage{graphicx}
\usepackage{times}
\usepackage{amsmath}
\usepackage{dcolumn}
\usepackage{latexsym,amsmath,amssymb,bm,euscript}
\begin{document}

\title{Monopole Quantum Numbers in the Staggered Flux Spin Liquid}

\author{Jason Alicea}
\affiliation{Department of Physics, California Institute of Technology, Pasadena, CA 91125}

\date{\today}

\begin{abstract}

Algebraic spin liquids, which are exotic gapless spin states preserving all microscopic symmetries, have been widely studied due to potential realizations in frustrated quantum magnets and the cuprates.  At low energies, such putative phases are described by quantum electrodynamics in $2+1$ dimensions.  While significant
progress has been made in understanding this nontrivial interacting 
field theory and the associated spin physics, one important issue
which has proved elusive is the quantum numbers carried by so-called 
monopole operators.  Here we address this issue in the 
``staggered-flux'' spin
liquid which may be relevant to the pseudogap regime in high-$T_c$. 
Employing general analytical arguments supported by simple numerics, we argue that proximate phases encoded in the monopole operators include the familiar Neel and valence bond solid orders, as well as other symmetry-breaking orders closely related to those previously explored in the monopole-free sector of the theory.  Surprisingly, we also find that one monopole operator carries trivial quantum numbers, and briefly discuss its possible implications.

\end{abstract}
\pacs{}

\maketitle


\section{Introduction}

When frustration or doping drives quantum fluctuations sufficiently strong to destroy symmetry-breaking order even at zero temperature, exotic ground states known as spin liquids emerge.  ``Algebraic spin liquids'' comprise one class in which the spins appear ``critical'', exhibiting gapless excitations and power-law correlations which, remarkably, can be unified for symmetry-unrelated observables such as magnetic and valence bond solid fluctuations.  This unification of naively unrelated correlations is a particularly intriguing feature, in part because it constitutes a ``smoking gun'' prediction for the detection of such phases.  

While the unambiguous experimental observation of a quantum spin liquid (either gapless, or the related topological variety) remains to be fulfilled, there are a number of candidate materials which may host such exotic ground states.  Recently the spin-1/2 kagome antiferromagnet \emph{herbertsmithite} has emerged 
as a prominent example,\cite{kagomeExpt1,kagomeExpt2,kagomeExpt3,kagomeDisorder1,kagomeDisorder2,kagomeNMR} and several gapless spin liquid proposals\cite{kagomeASLshort,kagomeASLlong,kagomeASLdisorder,kagomeAVL,kagomeFermiSurface}, as well as a more conventional valence bond solid phase\cite{kagomeVBS1,kagomeVBS2,kagomeVBSseries1,kagomeVBSseries2}, have been put forth for this material.  Furthermore, the cuprates have long been speculated to harbor physics connected to an algebraic spin liquid---the so-called ``staggered-flux'' state which we will focus on here---in the pseudogap regime of the phase diagram (for a recent comprehensive review, see Ref.\ \onlinecite{HighTcRMP}).  

On the theoretical end, our understanding of algebraic spin liquids has grown dramatically over the past several years.  Such states are conventionally formulated in terms of fermionic, charge-neutral ``spinon'' fields coupled to a U(1) gauge field, whose low-energy dynamics is described by compact quantum electrodynamics in $2+1$ dimensions (QED3).  Much effort has been focused on addressing two basic questions concerning these states.  First, can they be stable?  In more formal terms, is criticality in QED3 protected, or are there relevant perturbations allowed by symmetry which generically drive the system away from the critical fixed point?  And second, if algebraic spin liquids are stable, what are the measurable consequences for the spin system?  Both are nontrivial questions that require consideration of two classes of operators in QED3---those that conserve gauge flux such as spinon bilinears, and ``monopole operators'' that increment the gauge flux by discrete units of $2\pi$.

While QED3 is known to be a strongly interacting field theory which lacks a free quasi-particle description, the theory can nevertheless be controlled by generalizing to a large number $N$ of spinon fields and performing an analysis in powers of $1/N$.  Within such a large-$N$ approach, the answer to the first question has been rigorously shown to be `yes'---such phases can in principle be stable.\cite{U1stability}  In particular, despite some controversy concerning the relevance of monopoles, it has now been established that such operators are strongly irrelevant in the large-$N$ limit, their scaling dimension scaling linearly with $N$.\cite{BKW,U1stability}  

Significant progress has also been made in addressing the second question, particularly in the monopole-free sector.  The effective low-energy QED3 theory for algebraic spin liquids is known to possess much higher symmetry than that of the underlying microscopic spin Hamiltonian, leading to the remarkable unification of naively unrelated competing orders noted above.  Furthermore, the machinery of the projective symmetry group\cite{QuantumOrder} allows one to establish how correlations of flux-conserving operators in QED3 relate to physical observables such as Neel or valence bond solid correlations\cite{MikeSF}, and the large-$N$ analysis additionally provides quantitative predictions for the corresponding scaling dimensions\cite{RantnerWen}.  

The physical content of monopole operators in QED3, however, is much less understood.  Essentially, the difficulty here is that, due to gauge-invariance, determining monopole quantum numbers requires examination of full many-body spinon wavefunctions, rather than just a few low-energy single-particle states as suffices, say, for the spinon bilinears.  Although the monopoles are highly irrelevant in the large-$N$ limit, their scaling dimensions may become of order unity for realistic values of $N$ (\emph{e.g.}, $N = 4$ for the staggered-flux state), so understanding the competing orders encoded in these operators becomes an important and physically relevant issue.  Moreover, since monopoles are allowed perturbations in compact QED3 which can in principle destroy criticality for small enough $N$, one would like to identify the leading symmetry-allowed monopole operators.
Some progress on these issues has been made for gapless spin liquids on the triangular and kagome lattices\cite{FermVortSpin1,AVLlong,AVLonethird,kagomeAVL,kagomeASLlong}, though in the important staggered-flux state the physics encoded in the monopoles remains a mystery\cite{MikeSF,LeonSubir}.  The goal of this paper is to generalize the techniques employed earlier in the former cases to deduce the monopole quantum numbers for the staggered-flux state and reveal the competing orders encoded in this sector of the theory.

\subsection{Assumptions and Strategy}

Let us at the outset discuss the core assumptions on which our quantum-number analysis will be based.  First, we will assume that it is sufficient to study monopoles at the mean-field level.  That is, we will treat the flux added by a monopole operator as a static background ``felt'' by the spinons.  This is reasonable coming from the large-$N$ limit, where gauge fluctuations are strongly suppressed, and is in fact the standard approach adopted when discussing such flux insertions (see, \emph{e.g.}, Ref.\ \onlinecite{BKW}).  
The second, and more crucial, assumption we employ is that the quantum numbers for the leading monopoles (those with the slowest-decaying correlations) can be obtained from the difference in quantum numbers between the mean-field ground states with and without the flux insertion.  Put more physically, the leading monopole quantum numbers are taken to be the momentum, angular momentum, \emph{etc.}, imparted to the spinon ground states upon flux insertion.  

The latter is equivalent to assuming that 1.) the flux insertion is ``adiabatic'' in the sense that the fermionic spinons remain in their relative ground state everywhere between the initial and final state and 2.) no Berry phases are accumulated during this evolution.  The first point follows because if the fermions remain in their relative ground before and after the flux insertion, then this ought to be true everywhere in between as well.  Such an assumption is quite delicate given that the mean-field states we will study are gapless in the thermodynamic limit.  We will not attempt to justify this point rigorously, but we note that treating the problem in this way is in the same spirit as the conventional mean-field treatment of flux insertions mentioned above.  If invalid, then treating flux insertions as a static background in the first place may not be a very useful starting point for addressing this problem.  

Assuming no Berry phases is equally delicate.  It is worth mentioning that this assumption is known to break down in certain cases.  As an illustration, consider the following gauge theory on the square lattice,
\begin{eqnarray}
  H &=& H_f + H_{G},
  \\
  H_{f} &=& v\sum_{\bf r}(-1)^{r_x+r_y}c^\dagger_{{\bf r}\alpha} c_{{\bf r}\alpha}   
  \nonumber \\
  &-& t\sum_{\langle{\bf r r'}\rangle}[c^\dagger_{{\bf r}\alpha} c_{{\bf r'}\alpha}e^{-i A_{\bf r r'}} + h.c.],
  \\
  H_{G} &=& -K \sum_{\square}\cos(\Delta \times A) + \frac{h}{2}\sum_{\langle{\bf r r'}\rangle} E_{\bf r r'}^2,
  \label{HG}
\end{eqnarray}
where $c_{{\bf r}\uparrow/\downarrow}$ are spinful fermionic operators, the first sum in Eq.\ (\ref{HG}) represents a lattice curl summed over all plaquettes, and the divergence of the electric field $E_{\bf r r'}$ is constrained such that 
\begin{equation}
  (\Delta\cdot E)_{\bf r} = 1-c^\dagger_{{\bf r}\alpha} c_{{\bf r}\alpha}.
\end{equation}
The standard electric-magnetic duality can be applied in the limit $v/t\rightarrow \infty$,\cite{LeonMonopoles} in which case one obtains a pure gauge theory with $(\Delta\cdot E)_{\bf r} = (-1)^{r_x+r_y}$.  Such an analysis reveals that the leading monopole operators carry nontrivial quantum numbers as a consequence of Berry phase effects,\cite{LeonMonopoles} even though the quantum numbers of the fermions clearly can not change in this limit.  The root of these nontrivial quantum numbers can be traced to the fact that the electric field divergence changes sign between neighboring sites.  If one alternatively considered a pure gauge theory with vanishing electric field divergence, then no such Berry phases arise.  Since in the staggered-flux state of interest the physical Hilbert space has exactly one fermion per site and thus a vanishing electric field divergence, we believe that it is reasonable to suspect that Berry phases do not play a role there as well.

Given these assumptions, we will adopt the following strategy below.  First, we will give a quick overview of the $\pi$-flux and staggered flux states, deriving a low-energy mean-field Hamiltonian for these states as well as the symmetry properties for the continuum fields.  We will then consider $\pm 2\pi$ flux insertions, and in particular obtain the transformation properties for the four quasi-localized zero-modes which appear.  Armed with this information, we will follow closely the monopole study of Refs.\ \onlinecite{FermVortSpin1} and \onlinecite{AVLlong} and constrain the monopole quantum numbers as much as possible using various symmetry relations which must generically hold on physical states, such as two reflections yielding the identity.  The ambiguities that remain will be sorted out by appealing to general quantum number conservation and simple numerical diagonalization for systems with convenient geometries and gauge choices.  This will allow us to unambiguously determine the monopole quantum numbers, subject to the above assumptions.  We will then explore the competing orders encoded in the monopole operators, and close with a brief discussion of some outstanding questions.

\section{Preliminaries}

\subsection{Overview of $\pi$-flux and staggered-flux states}

Although we will ultimately be interested in exploring monopole quantum numbers in the staggered-flux state, we will use proximity to the $\pi$-flux state in our analysis and thus discuss both states here.  Consider, then, a square-lattice antiferromagnet with Hamiltonian
\begin{equation}
  H = J \sum_{\langle {\bf r r'}\rangle} {\bf S}_{\bf r}\cdot {\bf S}_{\bf r'}.
  \label{spinH}
\end{equation}
Mean field descriptions of the $\pi$-flux and staggered-flux states can be obtained from Eq.\ (\ref{spinH}) by first decomposing the spin operators in terms of slave fermions via
\begin{equation}
  {\bf S}_{\bf r} = \frac{1}{2}f^\dagger_{{\bf r}\alpha}{\bm \sigma}_{\alpha \beta} f_{{\bf r}\beta},
  \label{S}
\end{equation}
where ${\bm \sigma}$ is a vector of Pauli spin matrices and the fermions are constrained such that there is exactly one per site.  As discussed in Refs.\ \onlinecite{SU2gaugesymmetry1,SU2gaugesymmetry2,QuantumOrder}, there is an SU(2) gauge redundancy in this rewriting.  The resulting bi-quadratic fermion Hamiltonian can then be decoupled using a Hubbard-Stratonovich transformation, giving rise to a simple free-fermion Hamiltonian at the mean-field level of the form
\begin{equation}
  H_{MF} = -t \sum_{\langle {\bf r r'}\rangle} [f_{{\bf r}\alpha}^\dagger f_{{\bf r'}\alpha} e^{-i a_{\bf r r'}} + \text{h.c.}].
\end{equation}

The $\pi$-flux state corresponds to an ansatz where the fermions hop in a background of $\pi$ flux per plaquette; \emph{i.e.}, $a_{\bf r r'}$ is chosen such that $(\Delta\times a) = \pi$ around each square.  This state retains the full SU(2) gauge redundancy inherent in Eq.\ (\ref{S}).  As the name suggests, the staggered-flux state corresponds to an ansatz in which the fermions hop in flux which alternates in sign between adjacent plaquettes; \emph{i.e.}, $(\Delta\times a) = \pm \Phi$, where $\Phi$ is the flux magnitude.  Note that this ansatz reduces to the $\pi$-flux ansatz when $\Phi = \pi$ since $\pi$ flux and $-\pi$ flux are equivalent on the lattice.  In contrast to the $\pi$-flux state, there is only a U(1) gauge redundancy remaining here.  Note also that, despite appearances, staggering the flux does not break translation symmetry.  Rather, translation symmetry (and others) are realized nontrivially as a result of gauge redundancy---the operators transform under a projective symmetry group\cite{QuantumOrder}.  Both ansatzes in fact preserve all microscopic symmetries of the original spin Hamiltonian, namely, $x$ and $y$ translations $T_{x,y}$, $\pi/2$ rotations about plaquette centers $R_{\pi/2}$, $x$-reflection about square lattice sites $R_x$, time reversal $\mathcal{T}$, and SU(2) spin symmetry.  Notably, there is no symmetry leading to conservation of gauge flux, which is why monopole operators are in principle allowed perturbations.  

\subsection{Continuum Hamiltonian and symmetry transformations}

To derive a continuum Hamiltonian and deduce how the fields transform under the microscopic symmetries, we will now choose a gauge and set $e^{i a_{\bf r r'}} = 1$ on vertical links and 
$e^{i a_{\bf r r'}} = (-1)^{y}$ on horizontal links.  Although this corresponds to $\pi$ flux, the transformations for the staggered-flux state can still be readily obtained from this choice.  Furthermore, adopting this starting point yields the same continuum Hamiltonian as if we had chosen a staggered-flux pattern, up to irrelevant perturbations.\cite{MikeSF}  To obtain the spectrum we take a two-site unit cell and label unit cells by vectors ${\bf R} = n_x {\bf \hat{x}}+ 2n_y {\bf \hat{y}}$ ($n_{x,y}$ are integers) which point to sites on sublattice 1; sublattice 2 is located at ${\bf R} + {\bf \hat{y}}$.  We denote the spinon operators on the two sublattices by $f_{{\bf R}\alpha 1,2}$, where $\alpha$ labels spin.  The band structure is straightforward to evaluate, and at the Fermi level one finds two Dirac points at momenta $\pm {\bf Q}$, with ${\bf Q} = (\pi/2,\pi/2)$.  Focusing on low-energy excitations in the vicinity of these Dirac points, a continuum theory can be derived by expanding the lattice fermion operators as follows,
\begin{eqnarray}
  f_{{\bf R}\alpha 1} &\sim& e^{i ({\bf Q}\cdot {\bf R}+\pi/4)}[\psi_{\alpha R 1} + \psi_{\alpha R2}]   \nonumber \\
  &+& e^{-i ({\bf Q}\cdot {\bf R}+\pi/4)}[\psi_{\alpha L 1} - \psi_{\alpha L 2}]
  \label{f1}
  \\
  f_{{\bf R}\alpha 2} &\sim& e^{i ({\bf Q}\cdot {\bf R}+\pi/4)}[-\psi_{\alpha R 1} +\psi_{\alpha R2}]   \nonumber \\
  &+& e^{-i ({\bf Q}\cdot {\bf R}+\pi/4)}[\psi_{\alpha L 1} + \psi_{\alpha L 2}].
  \label{f2}
\end{eqnarray}
Here we have introduced four flavors of two-component Dirac fermions $\psi_{\alpha A}$, where $\alpha$ labels the spin and $A = R/L$ labels the node.  
We then obtain the continuum mean-field Hamiltonian
\begin{equation}
  {\mathcal H}_{MF} \sim \int_{\bf x} -i v\psi^\dagger [\partial_x\tau^x + \partial_y \tau^y] \psi,
\end{equation}
where $v \sim t$ is the Fermi velocity and $\tau^{a}_{jk}$ are Pauli matrices that contract with the Dirac indices.  

It is a straightforward exercise to deduce the transformation properties of continuum fields from Eqs.\ (\ref{f1}) and (\ref{f2}).  For either the $\pi$-flux or staggered-flux states, these can be realized as follows:
\begin{eqnarray}
  T_x &:& \psi \rightarrow -i \tau^x \sigma^y \mu^z [\psi^\dagger]^t
  \label{Tx} \\
  T_y &:& \psi \rightarrow i \tau^x \sigma^y \mu^x [\psi^\dagger]^t
  \\
  R_x &:& \psi \rightarrow -\mu^x \tau^y \psi
  \\
  R_{\pi/2} &:& \psi \rightarrow e^{-i \frac{\pi}{4}\tau^z} e^{i \frac{\pi}{4}\mu^y}i \mu^x \psi
  \\
  \mathcal{T} &:& \psi \rightarrow -i \mu^y \tau^z [\psi^\dagger]^t,
  \label{T}
\end{eqnarray}
where in addition to the spin and Dirac matrices we have introduced Pauli matrices $\mu^a_{AB}$ that contract with the node indices.
In the $\pi$-flux and staggered-flux cases, these transformations can be followed by an arbitrary SU(2) and U(1) gauge transformation, respectively.  For the former, it will prove useful to consider a particle-hole gauge transformation $\mathcal{C}_G$ which is an element of the SU(2) gauge group and transforms the lattice fermion operators as
\begin{eqnarray}
  f_{{\bf R}\alpha 1} &\rightarrow& e^{i \pi R_x} i\sigma^y_{\alpha\beta} 
  f_{{\bf R}\beta 1}^\dagger
  \\
  f_{{\bf R}\alpha 2} &\rightarrow& -e^{i \pi R_x} i\sigma^y_{\alpha\beta} 
  f_{{\bf R}\beta 2}^\dagger  .
\end{eqnarray}
It follows that for the continuum fields we have
\begin{eqnarray}
  \mathcal{C}_G &:& \psi \rightarrow \tau^x \sigma^y [\psi^\dagger]^t.
  \label{CG}
\end{eqnarray}
We stress that in the staggered-flux state $\mathcal{C}_G$ reverses the sign of the flux microscopically and therefore does not represent a valid gauge transformation there.  

\subsection{Flux insertion and zero-modes}

Next we discuss the sector of the theory with $\pm2\pi$ flux inserted over a large area compared to the lattice unit cell.  Treating the flux as a static background, the mean-field Hamiltonian then becomes
\begin{equation}
  {\mathcal H}_{MF,q} = \int_{\bf x} -i v\psi^\dagger [(\partial_x-ia^q_x)\tau^x + (\partial_y -ia^q_y)\tau^y] \psi.
\end{equation}
The vector potential is chosen such that $\nabla\times a^q = 2\pi q$, where $q = \pm 1$ is the monopole charge.  It is well known that the above Hamiltonian admits one quasi-localized zero-mode for each fermion flavor,\cite{Jackiw} four in this case.  These zero-modes can be obtained by replacing $\psi_{\alpha A}({\bf x}) \rightarrow \phi_{\alpha A,q}({\bf x}) d_{\alpha A,q}$, where $\phi_{\alpha A,q}({\bf x})$ is the quasi-localized wavefunction and $d_{\alpha A,q}$ annihilates the corresponding state.  Employing the Coulomb gauge, the wave functions are simply
\begin{eqnarray}
  \phi_{\alpha A,+} &\sim& \frac{1}{|{\bf x}|}\binom{1}{0}
  \\
  \phi_{Aa,-} &\sim& \frac{1}{|{\bf x}|} \binom{0}{1}.
\end{eqnarray}
It follows that the zero-mode operators $d_{\alpha A,q}$ transform in exactly the same way as $\psi_{\alpha A j}$, so the transformations can be read off from Eqs.\ (\ref{Tx}) through (\ref{T}) and (\ref{CG}).  For example, under reflections, we have $d_{\alpha R/L,q} \rightarrow i q d_{\alpha L/R,-q}$.

Since gauge-invariant states are half-filled, two of the four zero-modes must be filled in the ground states here.  Thus, it will be convenient to introduce the following short-hand notation:
\begin{eqnarray}
  D_{1,q} &=& d_{\uparrow R,q}d_{\downarrow R,q} + d_{\uparrow L,q} d_{\downarrow L,q}
  \label{D1} \\
  D_{2,q} &=& d_{\uparrow R,q}d_{\downarrow R,q} - d_{\uparrow L,q} d_{\downarrow L,q}
  \\
  D_{3,q} &=& d_{\uparrow R,q}d_{\downarrow L,q} - d_{\downarrow R,q} d_{\uparrow L,q}
  \\
  D_{4,q} &=& d_{\uparrow R,q} d_{\uparrow L,q}
  \\
  D_{5,q} &=& d_{\uparrow R,q}d_{\downarrow L,q} + d_{\downarrow R,q} d_{\uparrow L,q}
  \\
  D_{6,q} &=& -d_{\downarrow R,q} d_{\downarrow L,q}.
  \label{D6}
\end{eqnarray}
Of these, $D_{1,2,3}$ are spin-singlets, while $D_{4,5,6}$ are spin triplets.  The transformation properties of these operators under the microscopic symmetries, as well as the gauge transformation $\mathcal{C}_G$ in the case of the $\pi$-flux state, are given in Table \ref{Dtable}.  Note that $\mathcal{C}_G$ changes the sign of the monopole charge $q$, indicating that the states with $+2\pi$ flux and $-2\pi$ flux are not physically distinct in the $\pi$-flux case.  We will use this fact to infer which of the leading monopole operators have dominant amplitudes in the neighboring staggered-flux state in Sec.\ \ref{CompetingOrders}.  

\begin{table}
\caption{\label{Dtable} Transformation properties of the operators $D_{j,q}$ defined in Eqs.\ (\ref{D1}) through (\ref{D6}) which fill two of the four zero-modes in the presence of a $2\pi q$ flux insertion.  The gauge transformation $\mathcal{C}_G$ applies only in the $\pi$-flux state.}
\begin{ruledtabular} 
\begin{tabular}{c | c | c | c | c | c | c } 
  & $T_x$ & $T_y$ & $R_x$ & $R_{\pi/2}$ 
  & $\mathcal{T}$ & ${\mathcal C}_G$ 
  \\ \hline 
  $ D_{1,q} \rightarrow$ & $-D_{1,-q}^\dagger$ &
  $-D_{1,-q}^\dagger$ & $-D_{1,-q}$ & $i q D_{1,q}$ &
  $-D_{1,q}^\dagger$ & $D_{1,-q}^\dagger$ 
  \\ \hline
  $ D_{2,q} \rightarrow$ & $-D_{2,-q}^\dagger$ &
  $D_{2,-q}^\dagger$ & $D_{2,-q}$ & $i q D_{3,q}$ &
  $D_{2,q}^\dagger$ & $D_{2,-q}^\dagger$ 
  \\ \hline
  $ D_{3,q} \rightarrow$ & $D_{3,-q}^\dagger$ &
  $-D_{3,-q}^\dagger$ & $-D_{3,-q}$ & $i q D_{2,q}$ &
  $D_{3,q}^\dagger$ & $D_{3,-q}^\dagger$ 
  \\ \hline
  $ D_{4,q} \rightarrow$ & $-D_{6,-q}^\dagger$ &
  $-D_{6,-q}^\dagger$ & $D_{4,-q}$ & $-i q D_{4,q}$ &
  $-D_{4,q}^\dagger$ & $-D_{6,-q}^\dagger$ 
  \\ \hline
  $ D_{5,q} \rightarrow$
    & $-D_{5,-q}^\dagger$ 
  & $-D_{5,-q}^\dagger$ & 
  $D_{5,-q}$ & $-i q D_{5,q}$ & $-D_{5,q}^\dagger$ &
  $-D_{5,-q}^\dagger$ 
  \\ \hline
  $ D_{6,q} \rightarrow$
    & $-D_{4,-q}^\dagger$ 
  & $-D_{4,-q}^\dagger$ &
  $D_{6,-q}$ & $-i q D_{6,q}$ & $-D_{6,q}^\dagger$ &
  $-D_{4,-q}^\dagger$ 
\end{tabular} 
\end{ruledtabular} 
\end{table} 

We pause now to comment in greater detail on the subtlety with determining the staggered-flux monopole quantum numbers.  Naively, one might suspect that these can be inferred from the transformation properties of the zero-modes, which we have at hand.  Realizing the microscopic symmetries, however, generically requires gauge transformations, which leads to inherent ambiguities in how the fields transform.  In particular, for the staggered-flux case, there is an arbitrary overall U(1) phase in the transformations quoted in Table \ref{Dtable}, and a still greater ambiguity in the $\pi$-flux state due to its larger SU(2) gauge group.  But the monopole operators are gauge-invariant, so one must instead examine the symmetries of the full many-body wavefunctions, which are gauge invariant, rather than single-particle states.  In what follows we will first deduce the transformation properties of flux insertion operators $\Phi^\dagger_{j,q}$ which add $2\pi q$ flux to the ground state and fill two of the zero modes,
\begin{equation}
  \Phi^\dagger_{j,q} = D_{j,q}^\dagger|q \rangle\langle0|.
  \label{halfmonopole} 
\end{equation}
Here $|q\rangle$ represents the filled Dirac sea in the presence of $2\pi q$ flux with all four zero-modes empty and $|0\rangle$ is the ground state in the absence of a flux insertion.  The monopoles we will ultimately be interested in will be simply related to these objects.
Once we know the transformation properties of $\Phi^\dagger_{j,q}$ it will be trivial to read off the 
monopole quantum numbers.

\section{Quantum Number Determination}

\subsection{Symmetry relations}

As a first step, we will now constrain the quantum numbers of the operators $\Phi_{j,q}^\dagger$ defined above using various symmetry relations which must hold when acting on gauge-invariant states.  In particular, we will utilize the following,
\begin{eqnarray}
  (R_x)^2 &=& 1  
  \label{2reflections}
  \\
  T_xT_y &=& T_y T_x
  \label{TxTy}
  \\
  R_x T_y &=& T_y R_x
  \label{RxTy}
  \\
  T_y R_{\pi/2} &=& R_{\pi/2} T_x
  \label{TxyRelation}
\end{eqnarray}
Furthermore, all lattice symmetries must commute with time-reversal (when acting on gauge-invariant states).  

Quite generally, we expect the following transformations to hold,
\begin{eqnarray}
  T_{x,y} &:& |q\rangle\langle0| \rightarrow e^{i \varphi^{q}_{x,y}}[\prod_{\alpha A}d_{\alpha A,-q}^\dagger]|-q\rangle\langle0|
  \label{Txy}
  \\
  R_{x} &:& |q\rangle\langle0| \rightarrow e^{i \theta^{q}_{x}}|-q\rangle\langle0|
  \label{Rx}
  \\
  R_{\pi/2} &:& |q\rangle\langle0| \rightarrow e^{i \theta^{q}_{\pi/2}}|q\rangle\langle0|
  \label{rot}
  \\
  \mathcal{T} &:& |q\rangle\langle0| \rightarrow [\prod_{\alpha A}d_{\alpha A,q}^\dagger]|q\rangle\langle0|
  \\
  \mathcal{C}_G &:& |q\rangle\langle0| \rightarrow e^{i \theta^{q}_{G}}[\prod_{\alpha A}d_{\alpha A,-q}^\dagger]|-q\rangle\langle0|.
  \label{CG2}
\end{eqnarray}
The last transformation holds only for the $\pi$-flux state.  All phases introduced above are arbitrary at this point, but will be constrained once we impose symmetry relations on gauge-invariant states which have two of the zero-modes filled.  Moreover, since time-reversal is anti-unitary, we have chosen the phases of $|q\rangle$ such that no additional phase factor appears under this symmetry. 

Consider reflections first.  Equation (\ref{2reflections}) and commutation with time-reversal imply that $e^{i\theta^q_x} = s$, for some $q$-independent sign $s$.  The value of $s$ is insignificant, however, since we can always remove it by sending $|+\rangle \rightarrow s |+\rangle$.  Hence we will take
\begin{equation}
  e^{i\theta^q_x} = 1.
\end{equation}  
For translations, Eqs.\ (\ref{TxTy}) and (\ref{RxTy}), as well as commutation with time-reversal, yield
\begin{equation}
  e^{i \varphi^q_{x,y}} = s_{x,y},
\end{equation}
for some unknown signs $s_{x,y}$.  Similarly, Eq.\ (\ref{TxyRelation}) and commutation with time-reversal allow us to determine $\theta^q_{\pi/2}$ up to signs $s^q_{\pi/2}$:
\begin{eqnarray}
  e^{i \theta^q_{\pi/2}} &=& i s^q_{\pi/2},
  \\
  s^+_{\pi/2}s^-_{\pi/2} &=& -s_x s_y.
\end{eqnarray}

Let us turn now to the $\pi$-flux state, where the mean-field Hamiltonian is invariant under the particle-hole transformation $\mathcal{C}_G$ as well.  For the moment we will treat this operation like the other physical symmetries, which is merely a convenient trick for backing out the quantum numbers of interest for the staggered-flux state.  In particular, we will assert that $\mathcal{C}_G^2 = 1$ and that this particle-hole transformation commutes with the physical symmetries when acting on half-filled states.  This yields
\begin{equation}
  e^{i \theta^q_G} = s_G,
\end{equation}
for an undetermined sign $s_G$, and also gives the useful constraint
\begin{equation}
  s^+_{\pi/2} = - s^-_{\pi/2}.
\end{equation}
It follows from the last equation that
\begin{equation}
  s_x = s_y.
\end{equation}
Since the staggered-flux mean-field continuously connects to the $\pi$-flux ansatz, we will assume that the latter two constraints hold in the staggered-flux case as well.  (We could alternatively obtain this result using the numerics from the next section, without appealing to the $\pi$-flux state.)

To recap, in our study of the flux-insertion operators $\Phi^\dagger_{j,q}$ thus far, we have shown that symmetry relations highly constrain how these objects transform, and proximity to the $\pi$-flux state constrained these transformations even further.  All that remains to be determined are the signs $s_x$ and $s^+_{\pi/2}$ which appear under $x$-translations and $\pi/2$ rotations.  In the following section we argue that these can be obtained by employing general quantum number conservation arguments supported by simple numerical diagonalization.

\subsection{Numerical Diagonalization}

To determine the remaining signs $s_x$ and $s^+_{\pi/2}$, we will now discuss our numerical diagonalization study of the mean-field Hamiltonian with and without a flux insertion, and discuss a more intuitive quantum number conservation argument which is consistent with these numerics.  The basic idea behind our numerics is that we will judiciously choose the system geometry and gauge such that the symmetry under consideration can be realized without implementing a gauge transformation.  This is a crucial point, as only in this case can we avoid overall phase ambiguities that would otherwise appear in such a mean-field treatment.  Once the single-particle wavefunctions with and without a flux insertion are at hand, one can proceed to deduce the transformation properties of the corresponding many-body wavefunctions and, in turn, the flux-insertion operators $\Phi^\dagger_{j,q}$ by using the results of the previous section.  

Consider first $\pi/2$ rotations.  Here we diagonalize the mean-field Hamiltonian in a square $L$ by $L$ system with open boundary conditions and $L$ odd so that the system is invariant under $\pi/2$ rotations about the central plaquette's midpoint.  For all flux configurations we choose a rotationally symmetric gauge so that $\pi/2$ rotations are realized trivially.  We work in the $\pi$-flux ansatz for simplicity, though staggering the flux can easily be done and clearly does not change any of the results.  Flux is inserted over the few innermost ``rings'' of the system, and the ``zero-modes'' that appear quasi-localized around the $2\pi$ flux can be unambiguously identified by examining the spread of their wave functions.  (The ``zero-modes'' here are pushed away from zero energy due to finite-size effects; for each spin, one is pushed to higher energy while the other to lower energy.)  We consider a variety of system sizes, with up to on the order of 1000 lattice sites, and obtain consistent results in all cases examined.  (More details on these numerics can be found in Ref.\ \onlinecite{AVLlong}, which carried out a similar study on the triangular lattice.)

In particular, by considering the six ways of filling the zero-modes, we find numerically that there are four $-1$ and two $+1$ rotation eigenvalues for the operators $\Phi^\dagger_{j,+}$.  To then back out the sign $s^+_{\pi/2}$, we use Eq.\ (\ref{rot}) and Table \ref{Dtable} to show that these operators must have four $-s^+_{\pi/2}$ and two $+s^+_{\pi/2}$ rotation eigenvalues.  It immediately follows that
\begin{equation}
  s^+_{\pi/2} = 1.
\end{equation}

Actually, one can recover this result without resorting to numerics using the following argument.  Note first that the quantum numbers for each single-particle state must be identical for the two spin species.  Assume that as flux is inserted, no single-particle levels cross zero energy, as is typically the case in our observations.  The quantum numbers for the states below zero-energy are then conserved under flux insertion.  For simplicity, let us assume that the half-filled state $|0\rangle$ with no added flux carries trivial quantum numbers (which is by no means essential).  This implies that if for each spin the lower zero-mode (\emph{i.e.}, the one pushed downward in energy due to finite-size effects) has eigenvalue $e^{i \alpha_{\pi/2}}$ under rotation, then all other negative-energy states must have eigenvalue $e^{-i \alpha_{\pi/2}}$.  Denote the upper zero-mode eigenvalue for each spin by $e^{i\beta_{\pi/2}}$.  One can then easily show that under rotation, the operators $\Phi_{j,+}^\dagger$ must have one trivial eigenvalue, one eigenvalue $e^{2i(\beta_{\pi/2}-\alpha_{\pi/2})}$, and four eigenvalues $e^{i(\beta_{\pi/2}-\alpha_{\pi/2})}$.  The only consistent possibility is for $e^{i(\beta_{\pi/2}-\alpha_{\pi/2})} = -1$, which yields $s^+_{\pi/2} = 1$ as deduced from numerics.  

Deducing the sign $s_x$ is more delicate.  To this end we consider the composite operation $R_x T_x$, which is convenient since it does not change the sign of the flux inserted.  This combination does, however, require a particle-hole transformation, so we can not simply read off the eigenvalues of the half-filled states from numerics as we did for the rotations.  An argument similar to the one raised in the previous paragraph does nevertheless allow us to make progress.  As before, we consider a finite-size system where $R_x T_x$ is a well-defined symmetry.  A system with periodic boundary conditions along the $x$-direction and hard-wall along the $y$-direction is particularly convenient since one can then insert $2\pi$ flux without any difficulty.  To make the eigenvalues well-defined here, we must imagine this flux being inserted slowly so that we can monitor the wavefunction continuously during the evolution.  Assuming no zero-energy level crossings (this has been verified in most cases; see below), then there must be at least one half-filled state with two zero-modes filled that carries the same quantum numbers as the original half-filled ground state before the flux insertion.  In particular, both states must be spin singlets.  Now, using Eqs.\ (\ref{Txy}) and (\ref{Rx}) along with Table \ref{Dtable}, one can readily show that the spin singlet operators $\Phi_{1,2,3;q}^\dagger$ all have eigenvalue $s_x$ under $R_x T_x$.  So we conclude that
\begin{equation}
  s_x = 1.
\end{equation}

Although we have now fully determined the transformation properties of the flux insertion operators $\Phi_{j,q}^\dagger$, it will be useful to specialize to the $\pi$-flux state and deduce the sign $s_G$ that appears under the particle-hole transformation $\mathcal{C}_G$.  For this purpose we consider the combination $T_x \mathcal{C}_G$, which is a simple translation whose eigenvalues are easy to determine numerically.  As above, we consider an $L_x$ by $L_y$ system with periodic boundary conditions along the $x$-direction and hard-wall along the $y$-direction, and choose the Landau gauge for all flux configurations so that $T_x \mathcal{C}_G$ can be realized without a gauge transformation.  Flux insertions are placed uniformly over several consecutive rows midway between the hard walls.  We restrict ourselves to the case where $L_x/2$ is odd, since the ``zero-modes'' that appear in the presence of $2\pi$ flux can be unambiguously identified for such systems.  As in our analysis of rotations, we examine the six ways of filling the two zero-modes, and find numerically that there are four $-1$ and two $+1$ eigenvalues under $T_x \mathcal{C}_G$ for the operators $\Phi_{j,q}^\dagger$.  Using Eqs.\ (\ref{Txy}) and (\ref{CG2}) and Table \ref{Dtable}, one can also deduce from our earlier results that these operators must have four $s_G$ and two $-s_G$ eigenvalues under $T_x\mathcal{C}_G$, implying that
\begin{equation}
  s_G = -1.
\end{equation}
Note that we have confirmed here that typically there are indeed no zero-energy level crossings during flux insertion.  Moreover, the sign $s_G$ can be recovered without numerics using the same logic as we outlined for rotations, though we will not repeat the argument here.

The transformation properties for the flux-insertion operators $\Phi^\dagger_{j,q}$ under all symmetries are summarized in Table \ref{Ftable}.  

\subsection{Definition of monopole operators}

We will now define the monopole operators as follows,
\begin{eqnarray}
  M_1^\dagger &=& \Phi^\dagger_{1,+} + \Phi_{1,-}
  \label{M1}
  \\
  M_2^\dagger &=& \Phi^\dagger_{2,+} - \Phi_{2,-}
  \\
  M_3^\dagger &=& \Phi^\dagger_{3,+} - \Phi_{3,-}
  \\
  M_4^\dagger &=& \Phi^\dagger_{4,+} + \Phi_{6,-}
  \\
  M_5^\dagger &=& \Phi^\dagger_{5,+} + \Phi_{5,-}
  \\
  M_6^\dagger &=& \Phi^\dagger_{6,+} + \Phi_{4,-}.
  \label{M6}
\end{eqnarray}  
We have organized these ``ladder'' operators such that the monopoles add the same quantum numbers when acting on ground states within the $q = 0,\pm 1$ monopole charge sectors.  For instance, $M_4^\dagger$ adds $S^z = 1$ by filling two spin-up zero-modes when acting on $|0\rangle$ and by annihilating two spin-down zero-modes when acting on $D^\dagger_{6,-}|-\rangle$.  Furthermore, these operators have been defined so that they transform into one another under the emergent SU(4) symmetry enjoyed by the critical theory\cite{MikeSF}, implying that all six have the same scaling dimension.  Thus the various competing orders captured by the monopoles are unified, just as is the case for those encoded in the spinon bilinears whose correlations are enhanced by gauge fluctuations\cite{MikeSF}.  Again, this constitutes a highly nontrivial, and in principle verifiable, experimental prediction which we will elucidate further below.  

Before exploring the competing orders, we note that there is another important set of related operators that one should consider, which are the following composites involving the monopole charge operator $Q$,
\begin{equation}
  \mathcal{M}_j^\dagger = \{M_j^\dagger,Q\}.
\end{equation}
[Such operators effectively send $\Phi_{j,-} \rightarrow -\Phi_{j,-}$ in Eqs.\ (\ref{M1}) through (\ref{M6}).]  Our analysis thus far does not enable us to distinguish which of these two sets of operators dominates at the staggered-flux fixed point.  The following argument, however, suggests that both sets have the same scaling dimension.  Consider the current $J^\mu = \frac{1}{4\pi}\epsilon^{\mu\nu\rho}F_{\nu\rho}$, where $F_{\nu\rho}$ is the field-strength tensor.  The monopole charge operator is given by an integral over $J^0$:
\begin{equation}
  Q = \int dx dy J^0,
\end{equation}
which clearly yields an integer $q$ if there is $2\pi q$ flux present.  To all orders in $1/N$, $J^\mu$ scales like an inverse length squared,\cite{BKW} implying that $Q$ has zero scaling dimension.  Typically knowing the scaling dimension of two operators is not sufficient to determine the scaling dimension of the composite.  However, since $Q$ is not a local operator, but rather an integral of a charge density, the scaling dimensions for the composites $M_j^\dagger Q$ are additive.  Thus the scaling dimensions for $M_j$ and $\mathcal{M}_j$ should be equal.

\begin{table}
\caption{\label{Ftable} Transformation properties of the flux-insertion operators $\Phi^\dagger_{j,q}$.  The gauge transformation $\mathcal{C}_G$ applies only in the $\pi$-flux state.}
\begin{ruledtabular} 
\begin{tabular}{c | c | c | c | c | c | c } 
  & $T_x$ & $T_y$ & $R_x$ & $R_{\pi/2}$ 
  & $\mathcal{T}$ & ${\mathcal C}_G$ 
  \\ \hline 
  $ \Phi_{1,q}^\dagger \rightarrow$ & $-\Phi_{1,-q}^\dagger$ &
  $-\Phi_{1,-q}^\dagger$ & $-\Phi_{1,-q}^\dagger$ & $\Phi_{1,q}^\dagger$ &
  $-\Phi_{1,q}^\dagger$ & $-\Phi_{1,-q}^\dagger$ 
  \\ \hline
  $ \Phi_{2,q}^\dagger \rightarrow$ & $\Phi_{2,-q}^\dagger$ &
  $-\Phi_{2,-q}^\dagger$ & $\Phi_{2,-q}^\dagger$ & $\Phi_{3,q}^\dagger$ &
  $-\Phi_{2,q}^\dagger$ & $\Phi_{2,-q}^\dagger$ 
  \\ \hline
  $ \Phi_{3,q}^\dagger \rightarrow$ & $-\Phi_{3,-q}^\dagger$ &
  $\Phi_{3,-q}^\dagger$ & $-\Phi_{3,-q}^\dagger$ & $\Phi_{2,q}^\dagger$ &
  $-\Phi_{3,q}^\dagger$ & $\Phi_{3,-q}^\dagger$ 
  \\ \hline
  $ \Phi_{4,q}^\dagger \rightarrow$ & $-\Phi_{4,-q}^\dagger$ &
  $-\Phi_{4,-q}^\dagger$ & $\Phi_{4,-q}^\dagger$ & $-\Phi_{4,q}^\dagger$ &
  $-\Phi_{6,q}^\dagger$ & $\Phi_{4,-q}^\dagger$ 
  \\ \hline
  $ \Phi_{5,q}^\dagger \rightarrow$
    & $-\Phi_{5,-q}^\dagger$ 
  & $-\Phi_{5,-q}^\dagger$ & 
  $\Phi_{5,-q}^\dagger$ & $-\Phi_{5,q}^\dagger$ & $-\Phi_{5,q}^\dagger$ &
  $\Phi_{5,-q}^\dagger$ 
  \\ \hline
  $ \Phi_{6,q}^\dagger \rightarrow$
    & $-\Phi_{6,-q}^\dagger$ 
  & $-\Phi_{6,-q}^\dagger$ &
  $\Phi_{6,-q}^\dagger$ & $-\Phi_{6,q}^\dagger$ & $-\Phi_{4,q}^\dagger$ &
  $\Phi_{6,-q}^\dagger$ 
\end{tabular} 
\end{ruledtabular} 
\end{table}

\section{Competing Orders Encoded in Monopoles}
\label{CompetingOrders}

Now that we have all transformation properties for the flux-insertion operators $\Phi^\dagger_{j,q}$, we can finally deduce the quantum numbers of the six monopole operators defined in Eqs.\ (\ref{M1}) through (\ref{M6}) and explore the competing orders encoded in this sector of the theory.  
To this end, we will examine in detail the quantum numbers carried by the 12 Hermitian operators $M_{j}^\dagger + M_j$ and $i (M_j^\dagger-M_j)$.  These are summarized in Table \ref{HermitianMtable}, which is the main result of this paper.  (The quantum numbers carried by the Hermitian operators constructed from $\mathcal{M}_j$ can be trivially obtained from these, and we will only comment on such operators briefly at the end.)  

In contrast to the monopole scaling dimensions, the amplitudes for their correlations are non-universal and will only be related where required by symmetry.  We can gain some intuition for which operators have the dominant amplitudes, at least for weak staggering of the flux, by examining their quantum numbers under the particle-hole gauge transformation $\mathcal{C}_G$ in the $\pi$-flux ansatz.  Those which are even under this operation will survive projection into the physical Hilbert space, and are thus expected to have the largest amplitudes in the staggered-flux case as well.  Those which are odd vanish upon projection and should have suppressed amplitudes.  In passing we note that a similar analysis may provide useful, though non-universal, information for the flux-conserving operators as well.  The first six Hermitian monopole operators listed in Table \ref{HermitianMtable} are expected to have dominant amplitudes by the above logic, while the latter six should be suppressed.  We proceed now to discuss the results, comparing with previous results for the well-studied monopole-free sector\cite{MikeSF} where appropriate.  

\begin{table*}
\caption{\label{HermitianMtable} Quantum numbers carried by Hermitian monopole operators constructed from ${M}^\dagger_{j}$ defined in Eqs.\ (\ref{M1}) through (\ref{M6}).  In columns 3 through 5, we provide the eigenvalue if the operator is diagonal under the corresponding symmetry; otherwise the operator into which it transforms is given.  While all 12 operators have the same scaling dimension, the first six are expected to have the dominant amplitudes based on proximity to the $\pi$-flux state.}
\begin{ruledtabular} 
\begin{tabular}{c | c | c | c | c | c | c} 
  & Momentum $(k_x,k_y)$ & $R_x$ & $R_{\pi/2}$ 
  & $\mathcal{T}$ & Spin & Meaning 
  \\ \hline 
  $ i {M}_{1} + h.c.$ & 
  $(0,0)$ & $1$ & $1$ &
  $1$ & Singlet & Allowed perturbation
  \\ \hline 
  $ i{M}_{2} + h.c.$ & 
  $(0,\pi)$ & $1$ & $\rightarrow i{M}_{3} + h.c.$ &
  $1$ & Singlet &  VBS
  \\ \hline 
  $ i{M}_{3} + h.c.$ & 
  $(\pi,0)$ & $-1$ & $\rightarrow iM_2 + h.c.$ &
  $1$ & Singlet &  VBS
  \\ \hline 
  $ ({M}_{4}-M_6) + h.c.$ & 
  $(\pi,\pi)$ & $1$ & $-1$ &
  $-1$ & Triplet & Neel
  \\ \hline 
  $ {M}_{5} + h.c.$ & 
  $(\pi,\pi)$ & $1$ & $-1$ &
  $-1$ & Triplet & Neel
  \\ \hline 
  $ i(M_4+{M}_{6}) + h.c.$ & 
  $(\pi,\pi)$ & $1$ & $-1$ &
  $-1$ & Triplet & Neel
  \\ \hline \hline
  $ M_{1} + h.c.$ & 
  $(\pi,\pi)$ & $-1$ & $1$ &
  $-1$ & Singlet & $(\pi,\pi)$ component of scalar spin chirality
  \\ \hline 
  $ M_{2} + h.c.$ & 
  $(\pi,0)$ & $-1$ & $\rightarrow  M_3 + h.c.$ &
  $-1$ & Singlet & $(0,\pi)$ component of skyrmion density
  \\ \hline 
  $  {M}_{3} + h.c.$ & 
  $(0,\pi)$ & $1$ & $\rightarrow  M_2 + h.c.$ &
  $-1$ & Singlet & $(\pi,0)$ component of skyrmion density
  \\ \hline 
  $ i({M}_{4}-M_6) + h.c.$ & 
  $(0,0)$ & $-1$ & $-1$ &
  $1$ & Triplet & uniform vector spin chirality
  \\ \hline 
  $ i {M}_{5} + h.c.$ & 
  $(0,0)$ & $-1$ & $-1$ &
  $1$ & Triplet & uniform vector spin chirality
  \\ \hline 
  $ (M_4+{M}_{6}) + h.c.$ & 
  $(0,0)$ & $-1$ & $-1$ &
  $1$ & Triplet & uniform vector spin chirality
  
\end{tabular} 
\end{ruledtabular} 
\end{table*} 

The first operator in Table \ref{HermitianMtable}, interestingly, is  a singlet that carries \emph{no} nontrivial quantum numbers, and thus constitutes an allowed perturbation to the Hamiltonian; we discuss possible implications of this in the next section.  
Note that there is no symmetry-equivalent operator in the set of fermionic spinon bilinears, all of which carry nontrivial quantum numbers\cite{MikeSF}.  As an aside we comment that naively it may appear, given our quantum-number-conservation argument employed earlier, that having one singlet monopole operator carrying no quantum numbers is generic.  We stress that this is not the case.  We applied this argument in different geometries, which were designed so that the symmetry under consideration was realized in a particularly simple way.  Within each geometry, there must be one singlet flux insertion which transforms trivially as claimed.  But there are three such singlet operators, so the same one need not transform trivially in all cases.  Indeed, similar arguments applied to monopoles on the triangular lattice yield no such operators carrying trivial quantum numbers.\cite{AVLlong}

Remarkably, the next five operators encode perhaps the most natural phases for the square-lattice antiferromagnet---valence bond solid (VBS) and Neel orders.  We find it quite encouraging that these appear as the dominant nearby orders in our analysis.  Both VBS and Neel fluctuations are also captured by enhanced fermion bilinears, which are labeled $N_{C}^{1,2}$ and ${\bf N}_A^3$, respectively, in Ref.\ \onlinecite{MikeSF}.  It is intriguing to note that a recent study that neglected monopoles but took into account short-range fermion interactions found that the staggered-flux spin liquid may be unstable towards an SO(5)-symmetric fixed point, at which Neel and VBS correlations were unified.\cite{CenkeSO5}  
In light of our results, it would be interesting to revisit that work with the inclusion of monopoles, which for the physical value $N = 4$ may also play an important role.

The remaining six operators in the table are expected to have suppressed amplitudes compared to the operators discussed above.  The first of these transforms microscopically like
\begin{eqnarray}
  M_1 + h.c. &\sim& (-1)^{r_x+r_y}[{\bf S}_{a}\cdot({\bf S}_{b}\times {\bf S}_c) - {\bf S}_{b}\cdot({\bf S}_{c}\times {\bf S}_d)
  \nonumber \\
  &+& {\bf S}_{c}\cdot({\bf S}_{d}\times {\bf S}_a) - {\bf S}_{d}\cdot({\bf S}_{a}\times {\bf S}_b)  ],
\end{eqnarray}
where ${\bf S}_{a} = {\bf S}_{\bf r-\hat{y}}$, ${\bf S}_{b} = {\bf S}_{\bf r + \hat{x}}$, ${\bf S}_{c} = {\bf S}_{\bf r + \hat{y}}$, and ${\bf S}_{d} = {\bf S}_{\bf r - \hat{x}}$.  This operator  represents the $(\pi,\pi)$ component of the scalar spin chirality.  Apart from the finite momentum carried, $M_1 + h.c.$ carries the same quantum numbers as the enhanced fermion bilinear denoted $M$ in Ref.\ \onlinecite{MikeSF} that when added to the Hamiltonian drives the system into the Kalmeyer-Laughlin spin liquid\cite{KLshort,KLlong} which breaks time-reversal and reflection symmetry.

The next two singlet operators in the table transform like the following microscopic spin operators,
\begin{eqnarray}
  {M}_2 + h.c. &\sim& (-1)^{r_x}[{\bf S}_{1}\cdot({\bf S}_{2}\times {\bf S}_3) - {\bf S}_{2}\cdot({\bf S}_{3}\times {\bf S}_4) 
  \nonumber \\
  &+& {\bf S}_{3}\cdot({\bf S}_{4}\times {\bf S}_1) - {\bf S}_{4}\cdot({\bf S}_{1}\times {\bf S}_2) ]
  \\
  {M}_3 + h.c. &\sim& -(-1)^{r_y}[{\bf S}_{1}\cdot({\bf S}_{2}\times {\bf S}_3) - {\bf S}_{2}\cdot({\bf S}_{3}\times {\bf S}_4)
  \nonumber \\
  &+& {\bf S}_{3}\cdot({\bf S}_{4}\times {\bf S}_1) - {\bf S}_{4}\cdot({\bf S}_{1}\times {\bf S}_2)  ],
\end{eqnarray}
where we have used abbreviated notation with ${\bf S}_{1} = {\bf S}_{\bf r}$, ${\bf S}_{2} = {\bf S}_{\bf r + \hat{x}}$, ${\bf S}_{3} = {\bf S}_{\bf r + \hat{x} + \hat{y}}$, and ${\bf S}_{4} = {\bf S}_{\bf r + \hat{y}}$.  These monopole operators are closely related to an enhanced fermion bilinear, dubbed $N_C^3$ in Ref.\ \onlinecite{MikeSF}, that transforms like
\begin{eqnarray}
  N_C^3 &\sim& {\bf S}_{1}\cdot({\bf S}_{2}\times {\bf S}_3) - {\bf S}_{2}\cdot({\bf S}_{3}\times {\bf S}_4) 
  \nonumber \\
  &+& {\bf S}_{3}\cdot({\bf S}_{4}\times {\bf S}_1) - {\bf S}_{4}\cdot({\bf S}_{1}\times {\bf S}_2) .
\end{eqnarray}
Furthermore, Ref.\ \onlinecite{MikeSF} observed that $N_C^3$ also possesses the same symmetry as the $(\pi,\pi)$ component of the skyrmion density $\rho_S$,
\begin{equation}
  \rho_S = \frac{1}{4\pi} {\bf n}\cdot (\partial_x {\bf n}\times \partial_y{\bf n}),
\end{equation}
where ${\bf n}$ is a unit vector encoding slow variations in the Neel order parameter.  Consequently, ${M}_{2,3} + h.c.$ are symmetry equivalent to the $(0,\pi)$ and $(\pi,0)$ components of the skyrmion density.  

Finally, the last three triplets in the table transform like components of the spin operator
\begin{equation}
  {\bf S}_{1} \times {\bf S}_{3} - {\bf S}_{2} \times {\bf S}_{4}.
\end{equation}
Thus, these operators represent the uniform part of the vector spin chirality.  Enhanced fermion bilinears (${\bf N}_{A}^{1,2}$ in Ref.\ \onlinecite{MikeSF}) also represent vector spin chirality fluctuations, though at momenta $(0,\pi)$ and $(\pi,0)$.

What about Hermitian operators constructed from the composites $\mathcal{M}_j$?  Their quantum numbers can be easily deduced from those listed in Table \ref{HermitianMtable} by noting that the monopole charge is odd under translations, reflection, and $\mathcal{C}_G$ (in the $\pi$-flux state), but even under rotations and time-reversal.  Consequently, Hermitian $\mathcal{M}_j$ operators have relative momentum $(\pi,\pi)$ and opposite parity under reflection compared with the corresponding $M_j$ operators.  One can repeat the analysis given above for the latter, but we choose not to do so here.

\section{Discussion}

In this paper we have attempted to help resolve an outstanding issue in the study of algebraic spin liquids---namely, the quantum numbers carried by monopole operators---by considering the well-studied case of the staggered-flux state.  Our study builds on previous work\cite{FermVortSpin1,AVLlong} in the slightly different context of ``algebraic vortex liquids'', and can be generalized to other settings as well.  Essentially, our analysis was predicated on the assumption that the leading monopole quantum numbers can be deduced from the symmetry properties of the mean-field ground states with and without a flux insertion, with no additional Berry phase effects.  While we believe this is reasonable, and find the end results to be quite natural, such issues can be delicate since we are dealing with a gapless state.  Thus, we encourage further scrutiny of the conclusions reach in this paper.  Projected wavefunction studies of the type described in Ref.\ \onlinecite{kagomeASLinField} provide one distinct approach which can shed further light on the problem and may help to support our findings\cite{YingPC}.  A more dynamical treatment of monopoles, however, may ultimately be required.

Assuming we have succeeded in finding the quantum numbers of the leading monopole operators, one issue is worth discussing further.  Specifically, our analysis showed that there is one Hermitian monopole operator which carries no quantum numbers and thus represents a symmetry-allowed perturbation to the Hamiltonian.  An important issue is whether this perturbation destabilizes the staggered-flux state for the physical number of fermion flavors, which is $N = 4$.  The single-monopole scaling dimension computed in the large-$N$ limit in Ref.\ \onlinecite{BKW} is $\Delta_m \approx 0.265N$.  Extrapolating to $N = 4$ yields $\Delta_m \approx 1.06$, substantially lower than 3, suggesting that the symmetry-allowed monopole operator may constitute a relevant perturbation.  However, caution is warranted here (more so than usual in such extrapolations), since the subleading correction to the scaling dimension is generically an $N$-independent, possibly $O(1)$ number.  Given the obvious importance of this question for the high-$T_c$ problem, further studies of these scaling dimensions are certainly worthwhile.  And if the operator turns out to be relevant, what are the properties of the phase to which the system eventually flows?  An interesting possibility is that the system may flow off to a distinct spin liquid state, but it is also possible that dangerously irrelevant operators lead to broken symmetries.  This question is left for future work.

\acknowledgments{It is a pleasure to thank Leon Balents, Matthew P.\ A.\ Fisher, Michael Hermele, Anton Kapustin, Olexei Motrunich, Ying Ran, and Subir Sachdev for illuminating discussions.  This work was supported by the National Science Foundation through Grant DMR-0210790 and the Lee A.\ DuBridge Foundation.}


\end{document}